\documentclass[12pt,preprint]{aastex}

\shorttitle{Proper Motions of Radio Sources in Orion}
\shortauthors{G\'omez et al.}

\begin{document}

\title{Monitoring the Large
Proper Motions of Radio Sources in the Orion BN/KL Region}

\author{Laura G\'omez\altaffilmark{1}, 
Luis F. Rodr\'\i guez, Laurent Loinard, and Susana Lizano}
\affil{Centro de Radioastronom\'\i a y Astrof\'\i sica, UNAM,
Apdo. Postal 3-72,\\
Morelia, Michoac\'an, 58089 M\'exico\\
{l.gomez, l.rodriguez, l.loinard, s.lizano@astrosmo.unam.mx}}

\author{Christine Allen and Arcadio Poveda}
\affil{Instituto de Astronom\'\i a, UNAM,
Apdo. Postal 70-264, M\'exico, D. F., 04510 M\'exico
{chris@astroscu.unam.mx, poveda@servidor.unam.mx}}

\and

\author{Karl M. Menten}
\affil{Max-Planck-Institut f\"ur Radioastronomie, Auf 
dem H\"ugel 69, D-53121 Bonn, Germany
{kmenten@mpifr-bonn.mpg.de}}

\altaffiltext{1}{Max-Planck-Institut f\"ur Radioastronomie, Auf 
dem H\"ugel 69, D-53121 Bonn, Germany, E-mail: lgomez@mpifr-bonn.mpg.de}

\begin{abstract}
We present absolute astrometry of four radio sources 
in the Becklin-Neugebauer/Kleinman-Low (BN/KL) 
region, derived from archival data (taken in
1991, 1995, and 2000) as well as from new observations (taken in 2006). All
data consist of 3.6 cm continuum emission and were
taken with the Very Large Array in its highest
angular resolution A configuration. 
We confirm the large 
proper motions of the BN object, the radio source 
I (GMR I) and the radio counterpart of the infrared source n 
(Orion-n), with values from 15 to 26 km s$^{-1}$.
The three sources are receding from a point
between them from where they seem to have been ejected about
500 years ago, probably via the disintegration of a multiple
stellar system. We present simulations of very compact 
stellar groups that provide a plausible dynamical scenario
for the observations. The radio source Orion-n appeared as a double in the
first three epochs, but as single in 2006. We discuss 
this morphological change. The fourth source in the region, GMR D,
shows no statistically significant proper motions. We also present new,
accurate relative astrometry between BN and radio source I that
restrict possible dynamical scenarios for the region.
During the 2006 observations,
the radio source GMR A, located about $1{'}$ to the NW of the BN/KL region,
exhibited an increase in its flux density
of a factor of $\sim$3.5 over a timescale of one hour.
This rapid variability at cm wavelengths is similar to that previously found
during a flare at millimeter wavelengths that took place in 2003.

\end{abstract}

\keywords{astrometry --- ISM: individual (\objectname{Orion}) ---
radio continuum: stars --- stars: flare --- stars: pre-main sequence}

\section{Introduction}

The Orion BN/KL region is at the center of a remarkable, 
fast (30--100 km s$^{-1}$),
and massive ($\sim 10~M_\odot$) outflow,
with a kinetic energy of order
$4 \times 10^{47}$ ergs, which Kwan \& Scoville (1976)
ascribe to an explosive event and which also manifests itself as shock
excited molecular hydrogen (H$_2$) emission at near infrared
wavelengths (Beckwith et al. 1978).
Observations with higher angular resolution showed that
this molecular outflow is bipolar and weakly 
collimated, with its blueshifted lobe toward the northwest
and its redshifted lobe to the southeast
(Erickson et al. 1982; Rodr\'\i guez-Franco, Mart\'\i n-Pintado, \& 
Wilson 1999). This outflow was later resolved into ``fingers''
of H$_2$ emission, most probably
tracing shocked gas, that point away from the BN/KL region to the 
northwest and southeast (Allen \& Burton 1993; Stolovy et al. 1998; 
Schultz et al. 1999; Salas et al. 1999; Kaifu et al. 2000).
These observations indicate that a powerful ejection occurred 
in the BN/KL region recently. 
The H$_2$ ``finger'' system consists of over 100 individual bow shocks, 
which delineate a wide-angle bipolar outflow along a position angle
of about $135^\circ$. The proper motions of the H$_2$ knots suggest that
the explosion took place less than about 1,000 yr ago
(Lee \& Burton 2000; Doi et al. 2002; Nissen et al. 2007). 

The cause of this remarkable outflow remains unknown. 
Recently, Rodr\'\i guez et al. (2005) and G\'omez et al. (2005)
reported large proper motions (equivalent to velocities of the order of
a few tens of km s$^{-1}$) for the radio sources associated with the infrared sources
BN (the ``Becklin-Neugebauer object'')
and n, as well as for the radio source I. All three objects
are located at the core of the BN/KL region and appear 
to be moving away from a common center where they must all have been 
located about 500 years ago. This suggests that all three sources 
were originally part of a multiple massive stellar system that 
recently disintegrated as a result 
of a close dynamical interaction. Bally \& Zinnecker (2005) have suggested
that, given the uncertainty in the age of the explosion traced by
the high velocity gas that could have experienced deceleration, this 
phenomenon could have taken place simultaneously with  
the close dynamical interaction possibly traced by the proper motions
of the three radio sources. However, a detailed model that connects the two
events is still lacking.

In this paper we report new radio observations of the BN/KL region that
monitor and confirm the previously reported
large proper motions of source BN, I, and n.
In particular, we present very accurate relative astrometry between the sources
BN and I that restricts the past history of the motions of
these sources. We also report the flux variability of
objects in the zone. Finally, we propose a dynamical scenario that accounts for
the observed high velocity of the stellar objects.

\section{Observations and Data Reduction}

We have used 3.6 cm data from the VLA in its most extended A configuration 
to measure
the proper motions of radio sources in the Orion BN/KL
region. Observations of this region 
were available in the VLA archives for 1991
September 06, 1995 July 22, and 2000 November 13. We obtained
new observations on 2006 May 12.

The data were analyzed in the standard manner using the AIPS package
and the calibrated visibilities were imaged using weights intermediate
between natural and uniform (with the ROBUST parameter set to 0).
The data were also self-calibrated in phase and amplitude for
each epoch. To diminish the effects of extended emission, we used
only visibilities with baselines longer than 100 k$\lambda$,
suppressing the emission from structures larger than 2$''$. 

The VLA observations for 1995, 2000, and 2006
were all made with the same phase calibrator,
0541-056 (located at only $1\rlap.^\circ6$ from the core of Orion),
and this provides very reliable absolute astrometry. In contrast, the
observations made in 1991 September 06 used
as phase calibrator the source 0501-019
(located at $9\rlap.^\circ1$ from the core of Orion).
The astrometric error grows with the angular separation
between the phase calibrator and the source
(e. g. Pradel, Charlot, \& Lestrade 2006). To improve
the astrometry of the 1991 image, we used the 2000
positions of nine bright and compact radio sources in the core
of the Trapezium region, with no significant proper motions
in the study of G\'omez et al. (2005) to compare with the positions
in the 1991 image.
The average shift of the 2000 epoch positions with respect
to the 1991 epoch positions is
$\Delta \alpha = - 0\rlap.^s0007 \pm 0\rlap.^s0009$ and
$\Delta \delta = - 0\rlap.{''}035 \pm 0\rlap.{''}011$.
The shift in right ascension is not statistically significant but
we do find a systematic shift in declination between the
two epochs, implying that the 1991 declinations should be
corrected by $\Delta \delta = - 0\rlap.{''}035$.

\section{Results}

We have used these multiepoch VLA observations
taken over 15 years (1991-2006) to study the proper motions of
the four persistent and compact radio sources clearly detectable in the
BN/KL region (Tabs. 1 and 2 and Figs. 1 and 2). The positions of 
the sources at each epoch were determined using a linearized 
least-squares fit to a Gaussian ellipsoid function 
(task JMFIT of AIPS).

The source proper motions were then obtained by adjusting their
displacements over the celestial sphere with a linear fit (Tab.\ 2).
The proper motions of all four sources are consistent within 1-$\sigma$
with the results of G\'omez et al. (2005). At a distance of
414 pc (Menten et al. 2007), 1 $mas~yr^{-1}$ is equivalent
to 2.0 km s$^{-1}$ and the proper motions of BN, I, and n are in the
range of 15 to 26 km s$^{-1}$. We have thus confirmed the large
proper motions of the radio sources I, BN, and Orion-n found by
Rodr\'{\i}guez et al. (2005) and G\'omez et al. (2005).
The fourth source in the region, GMR D (Garay et al. 1987),
shows no statistically significant proper motions.
During the 2006 observations we detected a transient radio source
associated with the Orion G7 star Parenago 1839
(labeled in Fig. 2), that was not detected in the three previous
epochs and for which we cannot derive a proper motion. 
The absolute proper motions
of these sources are shown in Figure 2, after been corrected for the
mean absolute proper motions of 35 radio sources located in a region
with a radius of about 0.1 pc centered at the core of Orion
(G\'omez et al. 2005),
$\mu_{\alpha}~ cos \delta = +0.8 \pm 0.2~mas~yr^{-1};$
$\mu_{\delta} = -2.3 \pm 0.2 ~mas~yr^{-1}.$
Sandstrom et al. (2007) have noted that the mean
absolute proper motions of these 35 radio sources
differ at the 1-2 mas yr$^{-1}$ level from those given
by Kharchenko et al. (2005) from astrometry (made also with respect to 
the frame of remote quasars)
of 12 optical stars in the Orion cluster distributed over a region with
radius of 0.4 pc,
$\mu_{\alpha}~ cos \delta = +1.96 \pm 0.31~mas~yr^{-1};$
$\mu_{\delta} = -0.77 \pm 0.46 ~mas~yr^{-1}.$
At present it is unclear if these non-coincident values are due to
a real difference in the mean absolute proper motions of the
two populations of objects studied by G\'omez et al. (2005) and by 
Kharchenko et al. (2005), or to another cause.
We have used the values of G\'omez et al. (2005) that appear more appropriate
for embedded radio sources in Orion.

\section{A Dynamical Scenario for the Region}

The mass of BN is estimated to be  $8~ M_\odot < M_{BN} < 12.6~ M_\odot$
(Scoville et al. 1983, Rodr\'\i guez et al. 2005).  Since source $n$ is
moving with the largest proper motion, for simplicity we assume that
it has a smaller mass than sources BN and I and neglect its
contribution in the conservation of linear momentum and 
energy.  Using the observed proper motions in Table 2, conservation of
linear momentum along the direction of motion of BN implies that the mass of
source I is $M_I \simeq 1.5 \, M_{BN}$, i.e., $12~ M_\odot < M_I < 19~ M_\odot$.
We identify source I with a close binary formed by dynamical
interactions that has the negative binding energy of the system.
Energy conservation implies that the semimajor axis of the binary is
given by $a/{\rm AU} = 25 f (1-f)(M_{BN}/ 13~ M_\odot)$, where $f$ is the
mass fraction of the primary.  Thus, the maximum possible binary
separation, for $f=1/2$ (equal masses), is $a/{\rm AU} = 6 (M_{BN}/ 13~ M_\odot)$.
These estimates have taken into account the total velocities 
corrected for the radial velocities of each source (Rodr\'\i guez et al. 2005) that give
25 km s$^{-1}$ for BN and 15 km s$^{-1}$ for I.
Finally,
conservation of linear momentum in the direction perpendicular to the
motion of BN implies that the mass of source n is $M_n = 0.16 \, M_I$,
i.e., $2~ M_\odot < M_n < 3~ M_\odot$.

Using the masses estimated above, 
the kinetic energy involved in the three kinematically peculiar objects, BN, I
and n, is of order $10^{47} {\rm ergs}$.
To accelerate these objects to their
observed velocities from the typical small random motions of recently formed stars
(1--2 km s$^{-1}$),
one can invoke very close encounters in a multiple star system, as first
proposed by Poveda et al. (1967). We have recently updated these computations,
using the chain-regularization N-body code of Mikkola \& Aarseth (1993).  The results
illustrated in Poveda et al. (1967) were exactly 
reproduced by the new computations.

For the new examples, we 
simulated compact multiples composed of 5 stars of different masses (ranging
from 8 to 20 $M_\odot$), densely packed within radii of 400 AU ($\sim$0.0019 pc) and with a
velocity dispersion corresponding to the thermal velocity at a temperature of 10 K.
The stellar density required is thus $1.6 \times 10^{8}$ stars pc$^{-3}$, which appears to be
very large, since the largest stellar densities found observationally
in the Galaxy on scales of
0.01 -- 0.1 pc are in the range $1-4 \times 10^{5}$ stars pc$^{-3}$ 
(Figer et al. 2002; Megeath et al. 2005; Beuther et al. 2007). However, in our case
we are dealing with much smaller physical scales, of order 0.001 pc, and there is
at least one observational example of a very young stellar system 
where the required stellar densities are reached on these
small scales. In the surroundings of the protostar Cep A HW2 the results of
Curiel et al. (2002) and Mart\'\i n-Pintado et al. (2005) imply the presence of at least
four embedded young
stellar objects  within a projected area of $\sim 0\rlap.{''}8 \times 0\rlap.{''}8$ 
($600 \times 600~ AU^2$). If we assume that the physical depth of the region
equals its projected extent, we obtain a stellar density of
$1.6 \times 10^{8}$ stars pc$^{-3}$. Thus, the
initial conditions of our simulation, a small group of a few stars
in a region with dimensions of a few hundred AU, are consistent with the observations of
the Cep A HW2 region. An additional example of high stellar densities can be 
found in the  $\theta^1$ B Ori group.  This more evolved system is composed of five 
stars within about 1${''}$ of each other (Close et al. 2003), which, at 
the distance of Orion, corresponds to about 400 AU.  The stellar 
density in this group is similar to that of the Cep A HW2 
region. 

Preliminary results of the first 100 five-body cases fully confirm our
earlier findings.  We find that a sizable fraction of such compact
configurations produces one or more escapers with large
velocities (greater than about 30 km s$^{-1}$, i. e., 
runaway stars), after only about 2
crossing times.  The positive energy carried away 
by the high velocity escapers is
compensated by the formation of a tight binary or 
multiple. In over 70\% of the
cases the binary was composed of the two most 
massive stars.  To illustrate the
dynamical evolution of such compact multiple systems we plot in Figs. 3 and 4 
final positions and velocities (after 2.2 crossing times, corresponding to
about 650 years) for two five-body examples.  The similarity of these examples
to the observed configuration of the BN system (Fig. 2;  see also Fig. 3 in
G\'omez et al. 2005) is evident, if one takes into account 
the uncertainties in the estimated masses.
Nevertheless, these simulations are just an illustration of the physical 
process and explore a small range of the possible parameters.
A forthcoming paper will present results of many more N-body realizations of
several variants of the initial configurations
(Allen \& Poveda, in preparation). We emphasize that the initial
conditions we chose were not the result of 
``integrating backwards" the observed
positions and velocities of BN, I and n, but were taken to simulate
very dense stellar
conditions such as those observed in some small scale 
(hundreds of AUs)
regions of massive star formation.

\section{How close BN and source I were in the past?}

As we will see below, it is important to estimate how close were BN and source I in the past,
during their minimum separation
in the plane of the sky.
Fortunately, this minimum separation can be estimated accurately
using relative astrometry between BN and source I.

We define $x(t)$ and $y(t)$ as the separations in 
right ascension and declination of BN with respect to source I 
as a function of epoch $t$. If we assume that their proper motions are
linear, the displacements will be given by

$$x(t) = x(2000.0) + \mu_x~ (t-2000.0),$$

\noindent and

$$y(t) = y(2000.0) + \mu_y~ (t-2000.0),$$ 

\noindent where $x(2000.0)$ and $y(2000.0)$ are the displacements
in right ascension and declination for epoch 2000.0,
and $\mu_x$ and $\mu_y$ are the proper motions in 
right ascension and declination. These constants and their
errors can be determined very accurately
from the relative astrometry.

The separation of the sources as a function of time, $s(t)$, will then be
given by:

$$s(t) = (x^2(t) + y^2(t))^{1/2}.$$

Differentiating and equaling to 0, we find that
this minimum separation takes place at an epoch $t_{min}$ given by

$$t_{min} = -{{x(2000.0) \mu_x + y(2000.0) \mu_y} \over { \mu_x^2 + \mu_y^2}},$$ 

\noindent and that this minimum
separation is 

$$s_{min} = {{|x(2000.0) \mu_y - y(2000.0) \mu_x|} \over {(\mu_x^2 + \mu_y^2)^{1/2}}}.$$

The least squares fit to 14 data points from the VLA archive (shown in
Table 3 and Figure 5) gives
$x(2000.0) = -5.94 \pm 0.21$ arcsec, $\mu_x = -0.0125 \pm 0.0004$ arcsec yr$^{-1}$,
$y(2000.0) = 7.73 \pm 0.22$ arcsec, and $\mu_y = 0.0144 \pm 0.0004$ arcsec yr$^{-1}$.
These values give a total proper motion of $\mu_{tot} = 0.0191 \pm 0.0004$ arcsec yr$^{-1}$
at a P.A. of $-41\rlap.^\circ0 \pm 1\rlap.^\circ2$.
The proper motions are consistent within error, but more accurate, than those
reported by Tan (2004) and Rodr\'\i guez et al. (2005).
With these new values, we obtain:

$$t_{min} = 1490 \pm 11,$$

$$s_{min} = 0\rlap.{''}55 \pm 0\rlap.{''}16,$$

\noindent where the errors were calculated 
using standard propagation error theory (Wall \& Jenkins 2003).
We conclude that, about 500 years ago,
BN and source I were within $\sim$230$\pm$70 AU from each other
in the plane of the sky.

\section{Was BN ejected from $\theta^1$ C Ori?}

An alternative origin for the proper motions of BN has
been proposed by Tan (2004). In his scenario, the ejection of BN took
place from the $\theta^1$ C Ori binary system. While this model is attractive
because $\theta^1$ C Ori is a massive, hard, and eccentric binary as
required by such a 
dynamical event, our accurate relative astrometry seems to rule
out a close encounter between BN and $\theta^1$ C Ori in the past.

Tan (2004) made a comparison between the proper motions of BN (determined
relatively to source I), $\mu_{\alpha}~ cos \delta = -11.1 \pm 2.2~mas~yr^{-1};$
$\mu_{\delta} = +14.3 \pm 2.2 ~mas~yr^{-1}$, and the optical proper motions of
$\theta^1$ C Ori determined by van Altena et al. (1988),
$\mu_{\alpha}~ cos \delta = +1.4 \pm 0.17~mas~yr^{-1};$
$\mu_{\delta} = -1.8 \pm 0.16 ~mas~yr^{-1}$.
We note that the proper motions determined by
van Altena et al. (1988) are not in an absolute reference frame
(their system is defined by the average proper motions of their
reference stars) 
and the reliability of the comparison is uncertain.
Tian et al. (1996) have discussed the different reference frames for the
proper motions of stars in Orion obtained by five different
groups and find that they differ at the $\sim$1 mas yr$^{-1}$
level. Furthermore, none of these reference frames is absolute
in the sense of being referred to the remote quasars. 
In any case, for the sake of discussion,
we have followed the steps made by Tan (2004) to compare the
proper motions of BN and of $\theta^1$ C Ori and used
our relative astrometry for the
proper motions of BN with respect to source I
(consistent, but about 5 times
more accurate than those of Tan), as well as the proper
motions of
van Altena et al. (1988) for $\theta^1$ C Ori. Making the same analysis as
the one made in the previous section for the motions of BN with respect to source I, we obtain
$x(2000.0) = -35.2 \pm 0.2$ arcsec, $\mu_x = -0.0139 \pm 0.0004$ arcsec yr$^{-1}$,
$y(2000.0) = 60.2 \pm 0.2$ arcsec, and $\mu_y = 0.0162 \pm 0.0004$ arcsec yr$^{-1}$
for the relative proper motions between BN and $\theta^1$ C Ori.
With these values, we obtain:

$$t_{min} = -1216 \pm 62,$$

$$s_{min} = 12\rlap.{''}5 \pm 1\rlap.{''}0,$$

\noindent that is, the minimum separation between BN and $\theta^1$ C Ori took
place about 3,200 years ago, but this minimum separation was
large, more than 10-$\sigma$ away from
the close approach required in the model of Tan (2004).
These proper motions are shown in Figure 6.
This serious discrepancy can be alleviated if one is willing to
accept that the errors in the proper motions of $\theta^1$ C Ori
with respect to BN are larger. Adding in quadrature an
error of 1 mas yr$^{-1}$, the proper motions of $\theta^1$ C Ori
determined by van Altena et al. (1988)
become $\mu_{\alpha}~ cos \delta = +1.4 \pm 1.0~mas~yr^{-1};$
$\mu_{\delta} = -1.8 \pm 1.0 ~mas~yr^{-1}$.  
Under this assumption, the minimum separation
becomes $s_{min} = 12\rlap.{''}5 \pm 2\rlap.{''}4$,
about 5-$\sigma$ away from
the close approach required in the model of Tan (2004).
This may be more tolerable, but then the proper motion
of $\theta^1$ C Ori has a position angle with a large
error, PA = $142^\circ \pm 25^\circ$,
and the argument of the antiparallel alignment between the proper
motions of BN and $\theta^1$ C Ori is not compelling anymore.

One possible way to realign the proper motions of BN with respect
to $\theta^1$ C Ori in this scenario is to take into account that during
its close passage near source I, BN may have suffered a
gravitational deflection of several degrees (Tan 2004).
In any case, the possibility that a moving massive star passes close 
another massive star is very low.
BN appears projected at about 70$''$ from $\theta^1$ C Ori
in the plane of the sky. This corresponds to a physical separation
of $l \sim 4.3 \times 10^{17}$ cm. Within a sphere with this radius centered
on $\theta^1$ C Ori there are of the order of 10 massive stars.
This gives a local density of massive stars of
$N_* \sim 8.9 \times 10^{2}$ pc$^{-3}$.
The probability of an encounter between two stars at a
minimum separation of $R$ = 200 AU
(as derived from the relative astrometry between BN and source I),
is

$$p \simeq N_* \pi R^2 l \simeq 3.7 \times 10^{-4}.$$
  
\section{Possible Difficulties with the Cluster Disintegration Model}

Tan (2008) has pointed two possible difficulties with the model
of the disintegrating cluster originally located between BN and source I
proposed by Rodr\'\i guez et al. (2005) and G\'omez et al. (2005). 
The first difficulty is related with the fact that in some molecular tracers
(for example, the silicate line-of-sight extinction image of Gezari et al. 1998),
source I appears located close to the peak, suggesting it is stationary with
respect to the dense gas in the region. 

However, other tracers show a different
picture. The OH masers observed by 
Cohen et al. (2006; see their Figs. 1 and 8)
clearly are distributed between source I and BN.
These authors suggest that these masers trace shock fronts in a trail
produced by the passage of the moving sources.  
Part of the OH masers show a torus morphology approximately centered at
the position proposed by us to be the center of the BN, I, and n
ejection. 

It is also relevant that in many tracers
source I is not engulfed (this is, surrounded
in all directions) in dust and molecular gas
(Beuther et al. 2004; Wilson et al. 2000),
but that these emissions peak mostly to the SE of I.
This may be consistent with the motion of source I,
``clearing" molecular gas in its path to the SE.

The second possible difficulty pointed by Tan (2008) is related with 
the fact that if source I and source n are moving at velocities of
15 to 26 km s$^{-1}$, why are then they able to retain the gas associated
with them at the 100 AU scale?
In our interpretation, this gas is the result of
supersonic (hundreds of km s$^{-1}$) ejections from the
associated stars and
we do not expect it to be affected by their motion
at velocities an order of magnitude smaller with
respect to the surrounding medium.
However, if the new interpretation of Reid et al. (2007) of source
I as an ionized disk is correct, there is a problem in that it will
be hard to understand how a source that had a close encounter in the
recent past (500 years ago with source I) has managed to rebuild and retain
its circumstellar disk. 

Finally, we note that also BN has associated
ionized gas at the 100 AU scale. The observed deconvolved size
of BN at 3.6 cm is $\sim 0\rlap.{''}17 \times 0\rlap.{''}08$ 
(see Table 1)
or $\sim$70 AU $\times$ 30 AU at a distance of 414 pc. 
The explanation
for this gas is usually in terms of a hypercompact
HII region, although other possible explanations
have been discussed by Scoville et al. (1983). 
In Figure 7 we show a high angular resolution, high sensitivity
7 mm image of BN, produced from the same dataset that Reid et al. (2007) 
used to study the radio source I. 
The total flux density of the source at this wavelength is
26.4$\pm$0.7 mJy and its deconvolved dimensions are
$0\rlap.{''}075 \pm 0\rlap.{''}003 \times 0\rlap.{''}042\pm 0\rlap.{''}002; PA = 56^\circ \pm 2^\circ.$
The larger flux density and smaller angular dimensions observed at
7 mm with respect to 3.6 cm (see Table 1) are consistent with those
expected for a region of ionized in which the electron density
decreases as a function of radius (Reynolds 1986).
 
But, if BN is moving, why we do not see a clear ``tail"
of recombining gas as expected for a star that moves in a stationary medium
(Raga et al. 1997)?
The solution is probably in the very fast recombination time
of BN as compared with the time it takes BN to displace by a distance equal
to its own diameter.
Along the axis of motion, BN has an angular size of about 80 mas (30 AU
at a distance of 414 pc).
Moving at 10.8 mas yr$^{-1}$, we expect that the source will displace
a distance comparable with its size in a time of order
7 years. For the recombining ``tail" not to be detectable
we need for the ionized gas in BN to have a recombination time
much shorter than 7 years. From the 7 mm observations of Rodr\'\i guez et al.
(2008, in preparation) and from Moran et al. (1983) and Scoville et al. (1983),
we estimate an electron density of
$2 \times 10^7$ cm$^{-3}$, that implies a very fast recombination time of only
a few days. Then, we do not expect to see much of a recombining tail
as BN moves in the surrounding medium.

\section{Comments on Individual Sources in the BN/KL Region}

\subsection{Orion-n}

This infrared source, detected by
Lonsdale et al. (1982),
has recently been studied at mid-IR wavelengths
by Greenhill et al.
(2004a) who infer a luminosity of order 2,000 $L_\odot$ for it.
The associated radio source was first reported by Menten \& Reid (1995;
their source ``L"), who found it to be double in their VLA 3.6 cm image 
taken in 1994. Our data (see Fig. 8) shows that in the 1991, 1995, and 2000 images
the source has remained
double, with a north-south separation of about $0\rlap.{''}35$.
Remarkably, in the 2006 image
Orion-n appears as a single radio source. We believe that this morphological
change is real and not a consequence of different angular resolutions
since all the data analyzed have very similar angular resolution.
This result suggests that an alternative explanation could be
that the source has no real proper motions and that the apparent motion is a result
of the southern component becoming much brighter than the northern one over the 
time baseline of the observations. However, the positions fitted
(using the AIPS task JMFIT) to the 
1991, 1995, and 2000 images assuming that two Gaussians are present
(see Fig. 1) show a progressive shift in the position of the two components
whose average 
extrapolates well to the position of the single component used to fit
the 2006 image. 

The positions of the individual components of the double
radio source for 1991, 1995, and 2000
as well as that of the single source in 2006 are shown in Fig. 1.
The northern and
southern components of the double source
for the first three epochs have been fitted with dashed lines,
while all components have been fitted to a solid line.
This figure shows that the single source observed in 2006 is the
centroid of the double source seen in the other epochs
and not one of the components that became dominant.

Greenhill et al. (2004a) and
Shuping, Morris, \& Bally (2004) have analyzed
their mid-infrared images of Orion-n and conclude that it is slightly elongated
(at the arc sec scale) in the east-west direction, suggesting that it may trace
a disk-like structure approximately perpendicular to the axis joining the double
radio source. From submillimeter observations of
the dust emission from this source,
Beuther et al. (2004)
estimate a mass of about 0.27 $M_\odot$ for the associated gaseous
structure, a value consistent with that
expected for the disk of a massive
young star. Under this interpretation, the radio emission from source n would
be tracing an ionized outflow, or thermal jet. Although not frequent in
bipolar ionized outflows, the change from double-peaked to single-peaked source
and viceversa,
has been observed in a handful of sources (Cohen, Bieging, \& Schwartz
1982; Bieging, Cohen, \& Schwartz 1984; Rodr\'\i guez et al. 2007;
Loinard et al. 2007), and is possibly caused by the ejection of
clumps of ionized gas.
However, the reason for the morphological change observed in the radio counterpart
to Orion-n is unclear, and additional observations are required to
understand it. 

Despite the dramatic changes in its morphology, the total 3.6-cm flux 
density of source n does not appear to have undergone very
large changes over the four epochs of the observations, ranging from
1.5 to 2.2 mJy. The total 3.6 cm flux density measured in 1994 by Menten \& Reid
(1995), 2.0 mJy, also falls in this range.
The position shifts
observed in this source could potentially be due to morphological
changes. Nevertheless, the systematic displacement of the positions
over four epochs favors the more straightforward interpretation of true
proper motions.

\subsection{GMR I}

Greenhill et al.\ (2004b) and Reid et al. (2007), based on the
interpretation of the SiO and H$_2$O
emission observed around source I, proposed a model with a rotating disk and
a wide angle outflow pointing in the NE-SW direction. This model 
is proposed to replace that of 
Greenhill et al.\ (1998) in which the SiO masers were
proposed to form in the
limbs of a high velocity bi-conical outflow projected along a
NW-SE axis. Clearly, this source needs to be observed in even
greater detail to elucidate its nature and to determine its mass.

\subsection{BN}

Grosso et al. (2005) have discussed the presence of two X-ray sources near
BN: COUP 599a and COUP 599b. They conclude that neither source is BN itself,
but most probably deeply embedded
lower mass stars, either unrelated or companions to BN.

\subsection{GMR D}

This radio source is associated with an X-ray source
from the Chandra Orion Ultradeep Project Census
(Source COUP 662 in Grosso et al. 2005),
but it has no optical or near-infrared
counterpart. The association with X-ray emission suggests it
could be a T Tau star. It shows no statistically significant proper motions.
We confirm its known variability at radio wavelengths (e. g. Zapata et al. 2004), 
since we observe
its flux density to vary between 0.7 and 4.2 mJy
over the four epochs.

\subsection{The Transient Radio Source}

During the 2006 observations we detected an unresolved radio source
at position $\alpha(J2000) = 05^h~ 35^m~ 14\rlap.^s6565;$
$\delta(J2000) = -05^\circ~ 22'~ 33\rlap.{''}731$, with total flux density of 1.3$\pm$0.1 mJy. 
The radio source is located within $0\rlap.{''}1$
of the K7 star Parenago 1839 and we propose they are the same
object. The source is also associated within $1{''}$ with the
X-ray source COUP J053514.6-052233.
Unfortunately, the radio emission was not detected in the previous
three epochs and we do not have a proper motion for this source.
To our knowledge, this is the first radio detection of this source,
marked in Figure 2 as Parenago 1839.

\section{Conclusions}

We have confirmed the large proper motions, with values
from 15 to 26 km s$^{-1}$, found for the radio sources associated with
the BN object, the radio source I, and the infrared source Orion-n.
All three sources appear to be diverging from a point in between them, from where
they were apparently ejected about 500 years ago, probably via the disintegration of
a compact multiple stellar system.
We present simulations of the dynamical evolution of very compact groups of
stars that illustrate the physical process. These theoretical results imply that
these stars were part of a very dense young compact group
($\sim 10^8$ stars pc$^{-3}$), like those observed
today in small regions associated with Cep A HW2 and $\theta^1$ B Ori.

The radio source associated with the infrared source n shows a large
morphological change in the 2006 image, where after being observed as double
for many years, it appears as a single source. We believe that this
behavior is consistent with an interpretation in terms of a thermal jet.

Finally, in an Appendix we report a rapid centimeter flare during the 2006
observations in the source GMR A, of similar
characteristics than that observed at millimeter wavelengths in
2003.

\acknowledgments

CA thanks S. Aarseth for illuminating discussion about
his N-body codes. AP and CA thank A. Hern\'andez-Alc\'antara for
computational assistance.
LFR acknowledges the support from COECyT, Michoac\'an, M\'exico.
SL and LL are grateful to CONACyT, M\'exico and DGAPA, UNAM for
their support. The National Radio Astronomy Observatory (NRAO) is a
facility of the National Science Foundation operated under cooperative
agreement by Associated Universities, Inc.

\appendix

\section{Rapid Variability in GMR A}

The GMR A source is a strongly variable radio source (Garay et al. 1987; 
Felli 1993a, 1993b), a bright near-infrared point source (No. 573;
Hillenbrand \& Carpenter 2000), and a variable X-ray source (No. 297;
Feigelson et al.\ 2002). Bower et al.\ (2003a) reported the 
serendipitous discovery of a giant flare at 86 GHz from this radio
source with the Berkeley-Illinois-Maryland Array (BIMA) on 
2003 January 20, with a flux density that increased by a factor
of more than 5 on a
timescale of hours. The follow-up observations of Bower et al.
(2003b) with BIMA and the VLA and of
Furuya et al. (2003) with
the Nobeyama Millimeter Array showed that the source decayed within 
days of the outburst. These authors found GMR A to be highly
variable at 15 and 22 GHz on a timescale of hours to days during the period of the
mm flare, and the source is well known to be variable
at cm wavelengths, at least in the timescale of weeks to years
(Felli et al. 1993b; Zapata et al. 2004).

During the 2006 run, we found a rapid flux density increase by a
factor of $\sim$3.5 over a timescale of
about one hour (see Fig. 9). The flare observed at centimeter wavelengths
is remarkably similar to that observed previously in the millimeter.
The increase is in both cases in the order of a few and the rise timescale
is of order one hour. Furthermore, the circular polarization
(see Fig. 9) was in both cases of order (V/I) $\simeq$ $-$5 to $-$10\%
before the flare and, as emphasized by Bower et al. (2003b)
in the case of the millimeter flare,
no circular polarization is detected near the peak of the flare.  
We also note that the variations in the flux density of the source
are larger before than during the flare (see Fig. 8). 

This source is $\sim 1'$ to the NW of the phase center of the observations and
its astrometry is not as reliable as that of the four sources 
previously discussed.
We have assumed that the real error associated with its positions is
twice that given by the formal fit.
From our data, we derive proper motions of
$\mu_{\alpha} cos \delta = +4.8 \pm 1.8~mas~yr^{-1};$
$\mu_{\delta} = -1.6 \pm 2.1 ~mas~yr^{-1}$,
that are consistent at the 2-$\sigma$ level 
with the values reported
by G\'omez et al. (2005), Sandstrom et al. (2007), and Menten et al.
(2007). The values of Sandstrom et al. (2007) and  Menten et al.
(2007) were obtained from VLBI observations and should be very accurate.

A similar flare was reported recently by Forbrich et al. (2008) at 1.3 cm toward the
Orion radio source ORBS J053514.67-052211.2.
 



\clearpage

\begin{table}
\begin{center}
\small
\caption{Parameters of the 3.6 cm Sources in Orion BN/KL$^a$\label{tbl-1}}
\begin{tabular}{lcccc}
\tableline\tableline
 &\multicolumn{2}{c}{Position$^b$} & Total Flux
&  \\
\cline{2-3}
Source &  $\alpha$(J2000) & $\delta$(J2000) & Density (mJy) &
Deconvolved Angular Size$^c$  \\
\tableline
BN object & 05 35 14.1099 &  -05 22 22.741 & 4.8$\pm$0.1 &
$0\rlap.{''}17\pm0\rlap.{''}01\times 0\rlap.{''}08\pm0\rlap.{''}03;~+63^\circ
\pm 7^\circ$  \\%
Orion$-$n & 05 35 14.3546 & -05 22 32.776  & 2.2$\pm$0.2 &
$0\rlap.{''}50\pm 0\rlap.{''}04 \times \leq 0\rlap.{''}09 ;~+20^\circ
\pm 3^\circ$  \\%
GMR I   & 05 35 14.5141  & -05 22 30.556  &  1.2$\pm$0.1 &
$0\rlap.{''}19\pm 0\rlap.{''}05 \times \leq 0\rlap.{''}15 ;~+136^\circ
\pm 23^\circ$  \\%
GMR D  & 05 35 14.8963  & -05 22 25.390  & 0.7$\pm$0.1 &
$\leq 0\rlap.{''}20$  \\%
\tableline
\end{tabular}
\tablenotetext{a}{The positions, flux densities and deconvolved sizes reported here are from 
the May 12, 2006 observations.}
\tablenotetext{b}{Units of right
ascension are hours, minutes, and seconds
and units of declination are degrees, arcminutes, and arcseconds. Positional accuracy
is estimated to be $\sim 0\rlap.{''}01$.}
\tablenotetext{c}{Major axis $\times$ minor axis; position angle of major axis.}
\end{center}
\end{table}

\clearpage

\begin{deluxetable}{lrrrr}
  \tablewidth{13.5cm}
\tablecaption{Absolute Proper Motions of the Radio Sources\tablenotemark{a}}
  \tablehead{ 
\colhead{} & 
\colhead{$\mu_{\alpha} \cos \delta$} &
\colhead{$\mu_{\delta}$} &
\colhead{$\mu_{total}$} &
\colhead{P.A.} \\%
\colhead{Source} &
\colhead{(mas yr$^{-1}$)} & 
\colhead{(mas yr$^{-1}$)} & 
\colhead{(mas yr$^{-1}$)} & 
\colhead{($^\circ$)} \\%
  } \startdata

BN object & $-$5.3  $\pm$ 0.9 &    9.4 $\pm$ 1.1  & 10.8  $\pm$ 1.0  & $-$29 $\pm$   5  \\%
Orion$-$n\tablenotemark{b}      &    0.0 $\pm$ 0.9 & $-$13.0  $\pm$ 1.2 & 13.0 $\pm$ 1.2 & 180 $\pm$ 4\\%
GMR I   &  4.5  $\pm$ 1.2 & $-$5.7 $\pm$ 1.3  &  7.3  $\pm$ 1.2  &   142 $\pm$   10  \\%
GMR D    & $-$0.4  $\pm$ 1.3 & $-$1.8 $\pm$ 1.6  &  1.8  $\pm$ 1.6  &$-$166 $\pm$  42  \\%

  \enddata 	 	     
\tablenotetext{a }{The errors quoted in this Table are 1--$\sigma$. At a distance of
414 pc (Menten et al. 2007), 1 $mas~yr^{-1}$ is equivalent
to 2.0 km s$^{-1}$.}
\tablenotetext{b }{See text for the proper motions reported here.}
\end{deluxetable}

\clearpage

\begin{deluxetable}{lccccc}
  \tablewidth{18.5cm}
\tablecaption{VLA Data Used for the Determination of the Relative Proper Motions
between BN and source I}
\small
\tablehead{
\colhead{}  & \colhead{}  & \colhead{$\lambda$} 
& \colhead{Synthesized Beam} & $\Delta \alpha$\tablenotemark{b} & 
$\Delta \delta$\tablenotemark{b} \\  
\colhead{Epoch}  & \colhead{Project} & \colhead{(cm)}
& \colhead{($\theta_M \times \theta_m; PA$)\tablenotemark{a}} & (seconds) & (arcsecs) \\
}
\startdata
1985 Jan 19 (1985.04) & AM143 & 1.3 & $0\rlap.{''}095\times0\rlap.{''}088;-7^\circ$ &
-0.38540$\pm$0.00079 & 7.5123$\pm$0.0128  \\
1985 Jan 19 (1985.04) & AM143 & 2.0 & $0\rlap.{''}159\times0\rlap.{''}131;+32^\circ$ & 
-0.38550$\pm$0.00074 & 7.5105$\pm$0.0110  \\
1986 Apr 28 (1986.32) & AC146 & 2.0 & $0\rlap.{''}154\times0\rlap.{''}144;+10^\circ$ &
-0.38678$\pm$0.00096 & 7.5202$\pm$0.0133  \\
1991 Sep 02 (1991.67) & AM335 & 1.3 & $0\rlap.{''}104\times0\rlap.{''}095;-16^\circ$&
-0.39143$\pm$0.00074 & 7.5919$\pm$0.0110  \\
1994 Apr 29 (1994.33) & AM442 & 3.6 & $0\rlap.{''}227\times0\rlap.{''}198;+1^\circ$ &
-0.39408$\pm$0.00070 & 7.6425$\pm$0.0108  \\
1995 Jul 22 (1995.56) & AM494 & 3.6 & $0\rlap.{''}262\times0\rlap.{''}320;+32^\circ$ &
-0.39422$\pm$0.00079 & 7.6768$\pm$0.0116  \\
1996 Nov 21 (1996.89) & AM543 & 3.6 & $0\rlap.{''}324\times0\rlap.{''}253;+0^\circ$ &
-0.39541$\pm$0.00073 & 7.6988$\pm$0.0117  \\
1997 Jan 11 (1997.03) & AM543 & 3.6 & $0\rlap.{''}329\times0\rlap.{''}246;-10^\circ$ &
-0.39471$\pm$0.00078 & 7.7066$\pm$0.0122  \\
2000 Nov 10 (2000.86) & AM668 & 0.7 & $0\rlap.{''}059\times0\rlap.{''}046;-25^\circ$ &
-0.39839$\pm$0.00068 & 7.7309$\pm$0.0102  \\
2000 Nov 13 (2000.87) & AM668 & 3.6 & $0\rlap.{''}244\times0\rlap.{''}223;+3^\circ$ &
-0.39898$\pm$0.00078 & 7.7282$\pm$0.0121  \\
2002 Mar 31 (2002.25) & AG622 & 0.7 & $0\rlap.{''}047\times0\rlap.{''}025;+25^\circ$ &
-0.40002$\pm$0.00079 & 7.7541$\pm$0.0131  \\
2004 Nov 06 (2004.85) & AB1135 & 3.6 & $0\rlap.{''}229\times0\rlap.{''}202;+1^\circ$ &
-0.40249$\pm$0.00077 & 7.8079$\pm$0.0120  \\
2006 May 12 (2006.36) & AR593 & 3.6 & $0\rlap.{''}262\times0\rlap.{''}217;-2^\circ$ &
-0.40426$\pm$0.00077 & 7.8142$\pm$0.0119  \\
2007 Dec 14 (2007.95) & AR635 & 0.7 & $0\rlap.{''}213\times0\rlap.{''}180;+20^\circ$ &
-0.40410$\pm$0.00069 & 7.8365$\pm$0.0105  \\

\enddata
\tablenotetext{a}{Major axis$\times$minor axis in arcsec; PA in degrees}
\tablenotetext{b}{Positional offsets of BN with respect to source I in
right ascension and declination.}

\end{deluxetable}

\clearpage

\begin{figure}
\epsscale{0.7}
\plotone{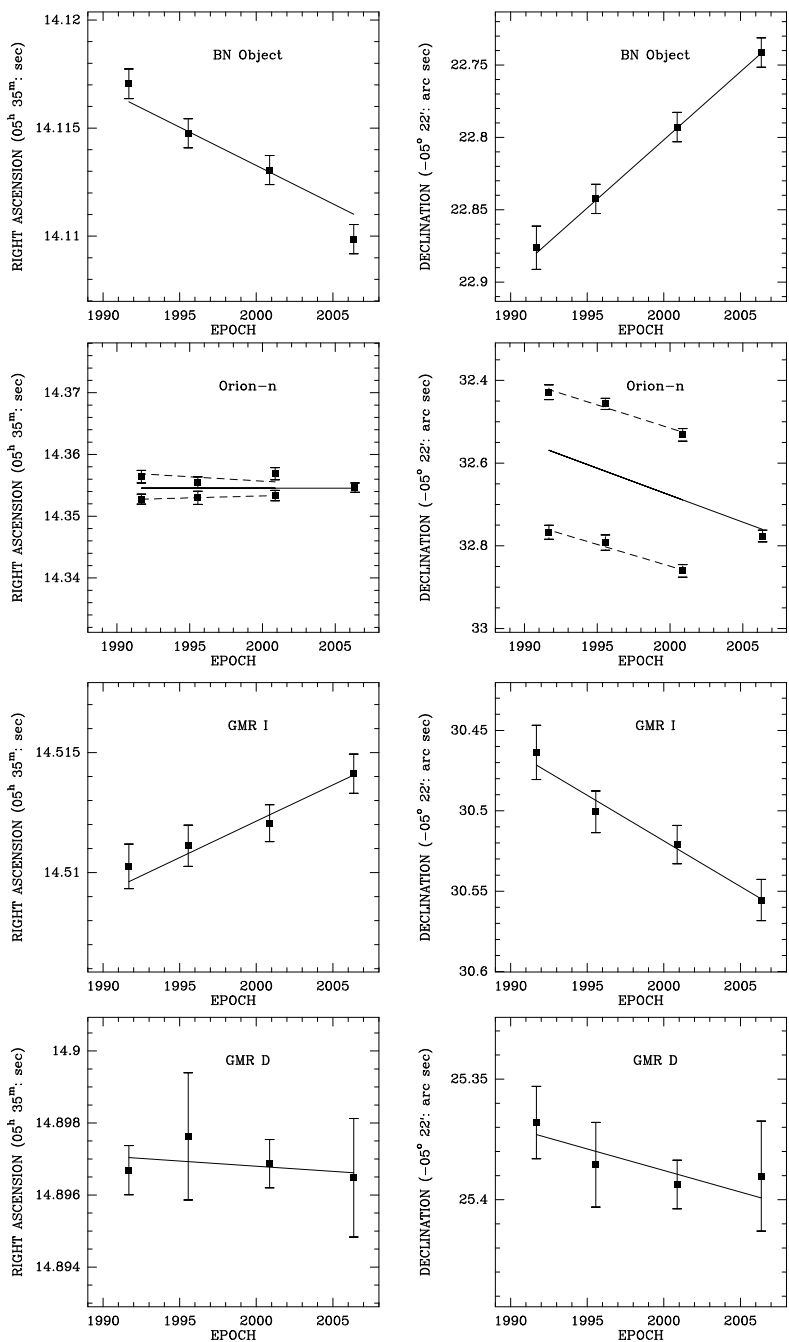}
\caption{Proper motions for the four persistent
radio sources in the Orion BN/KL region.
The solid lines are the least-squares fits to the data.
The proper motions of source Orion-n are described in the text.
\label{fig1}}
\end{figure}

\clearpage

\begin{figure}
\plotone{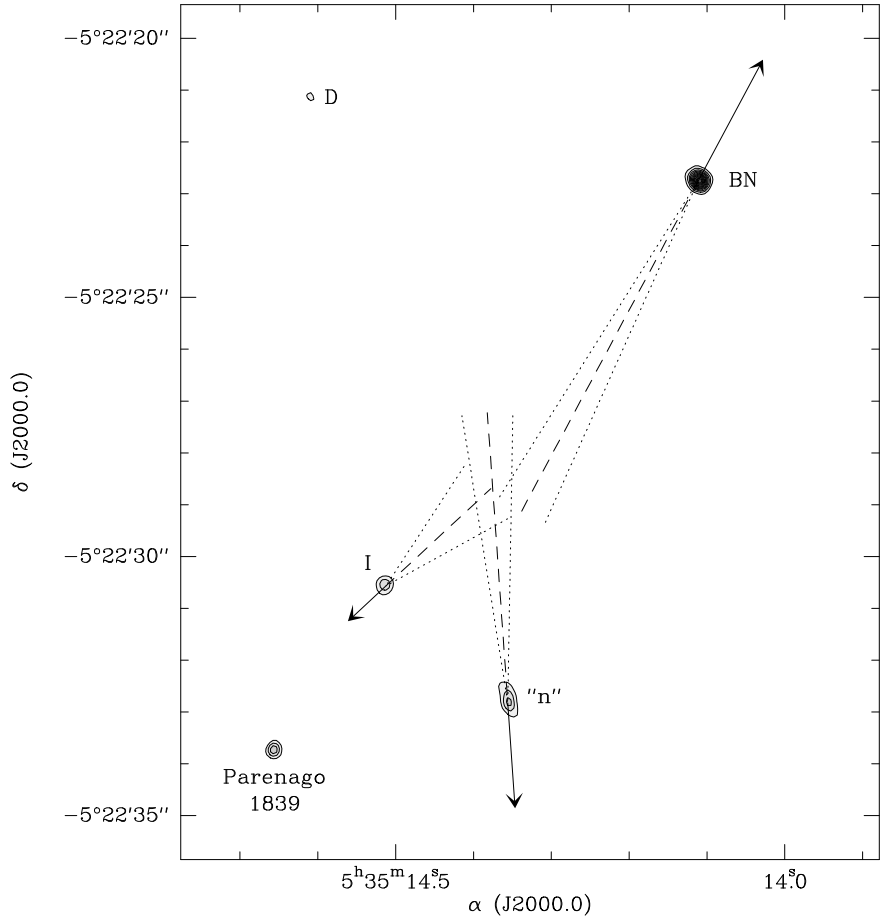}
\caption{VLA contour image at 8.46 GHz toward
the Orion BN/KL region 
for epoch 2006.36.
The first contour is 300
$\mu$Jy beam$^{-1}$ and increments are in units
of 150 $\mu$Jy beam$^{-1}$.
The angular resolution of the image is
$0\rlap.{''}26 \times 0\rlap.{''}22$; PA = $-2^\circ$.
The individual radio sources are identified.
The arrows indicate the direction and proper motion displacement for
200 years, in the rest frame of the Orion radio sources (G\'omez
et al. 2005). The dashed angles indicate the error in the
position angles of the proper motions.
At a distance of
414 pc (Menten et al. 2007), 1 $mas~yr^{-1}$ is equivalent
to 2.0 km s$^{-1}$
\label{fig2}}
\end{figure}

\clearpage

\begin{figure}
\epsscale{0.6}
\plotone{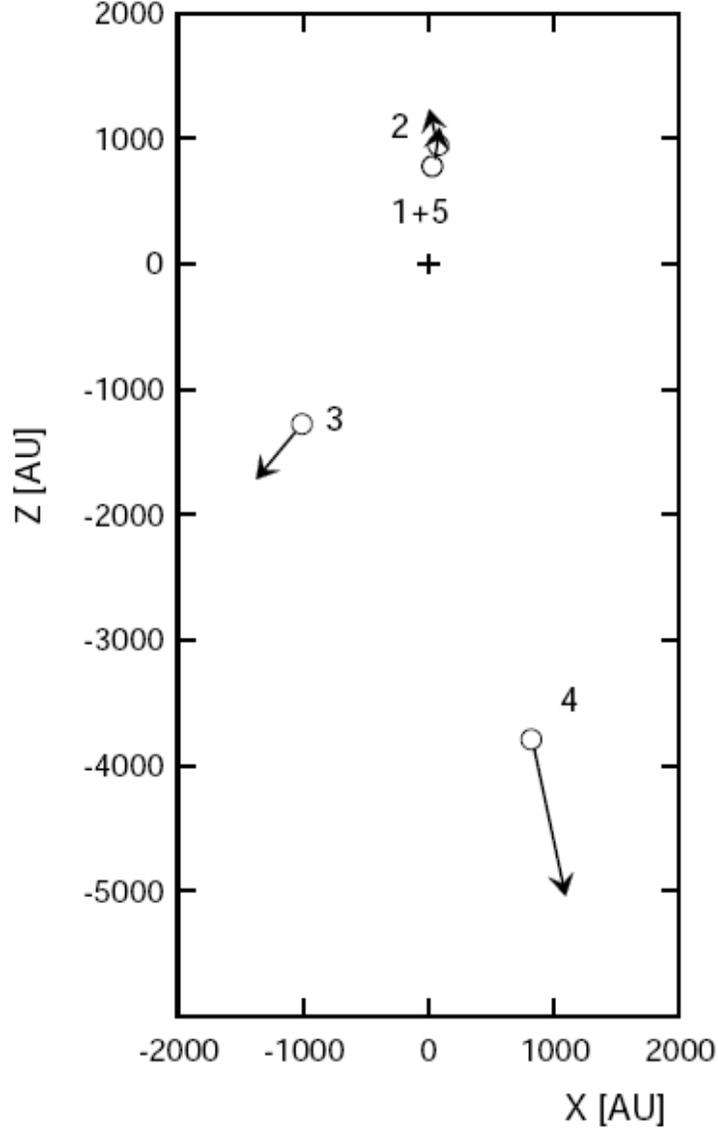}
\caption{Example 1.  Positions and velocities on the plane XZ for 5 stars
after 2.2 crossing times (650 years). The center 
of mass in marked by a cross.  The Y component of the velocity vector
of the runaway star (star 4) is small compared to 
the XZ components.  The space
velocities of  the BN-I system lie mostly on the plane of the sky and are
observed as transverse velocities. Thus, this figure is directly comparable to
Figure 2 of the BN system in the present paper. In this 
example  $V_{xz}(4)$ = 40 km s$^{-1}$, $V_{xz}(1+5)$ = 8.4 km s$^{-1}$,
the major semiaxis of binary $a$(1+5) = 13.6 AU, 
and the binding energy of this binary is
$E$(1+5)= $-2 \times 10^{47}$ ergs. The individual masses are $M(1) = M(2) =
16 M_\odot$, $M(3) = M(4) = 8M_\odot$, $M(5) = 20 M_\odot$. The total kinetic energy of 
stars 4, 2 and 3 plus that of the
center of mass of the binary (1+5) is $1.9 \times 10^{47}$ 
ergs.  The runaway star (star 4)
has reached a projected distance of 2879 AU from the center of mass, and the
binary (1+5) a projected distance of 704 AU.
\label{fig3}}
\end{figure}

\clearpage

\begin{figure}
\epsscale{0.8}
\plotone{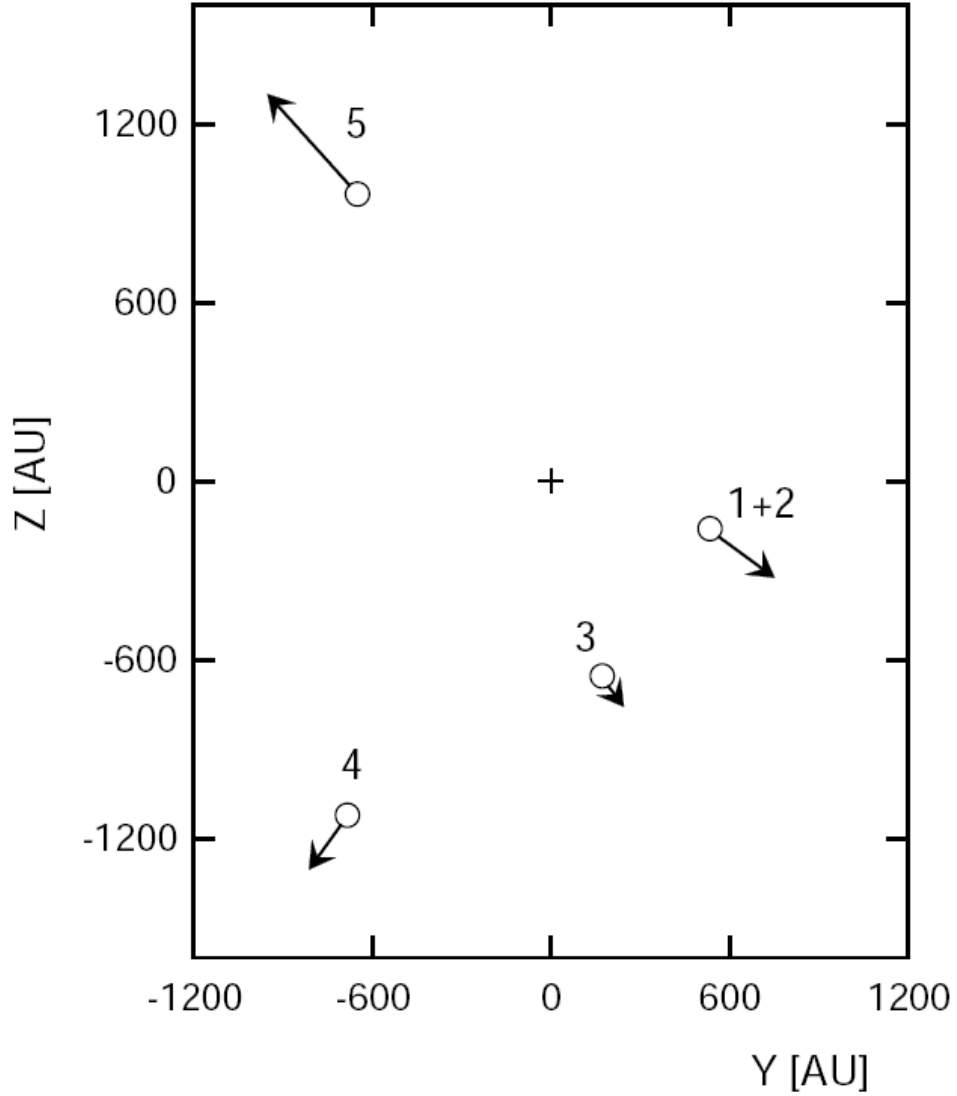}
\caption{Example 99.  Similar to Figure 3, but on the plane YZ.  In this
example, $V_{yz}(5)$ = 30 km s$^{-1}$, $V_{yz}(1+2)$ = 18.7 km s$^{=1}$, $a$(1+2) = 7.6 AU,  
$E$(1+2) = $-3 \times 10^{47}$ ergs, and the total kinetic energy of stars 3,4 and 5 plus that of
the center of mass of the binary (1+2) is $2.6 \times 10^{47}$ ergs. The runaway star
(star 5) has reached a projected distance of 1163 
AU from the center of mass, and the binary (1+2)
a projected distance of 557 AU.
\label{fig4}}
\end{figure}

\clearpage

\begin{figure}
\epsscale{0.5}
\plotone{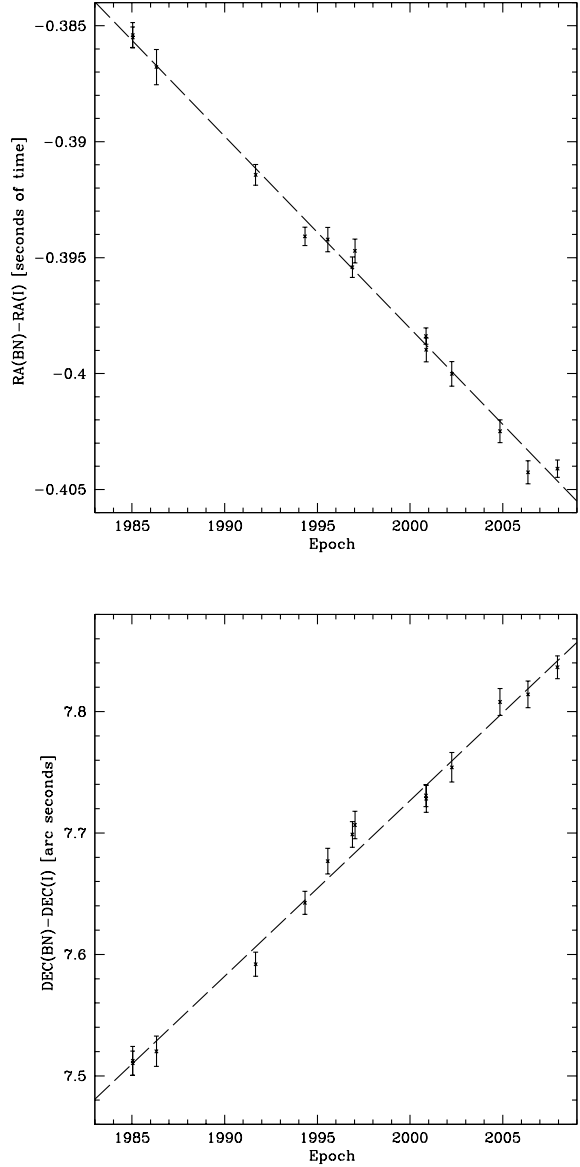}
\caption{Relative proper motions of BN with respect to source I.
The data used for this graph is listed in Table 3. The dashed line
is the weighted least-squares fit to the data points.
The relative proper motions in right ascension and
declination are $\mu_x = -0.0125 \pm 0.0004$ arcsec yr$^{-1}$
and $\mu_y = 0.0144 \pm 0.0004$ arcsec yr$^{-1}$, respectively.
\label{fig5}}
\end{figure}

\clearpage

\begin{figure}
\epsscale{1.0}
\plotone{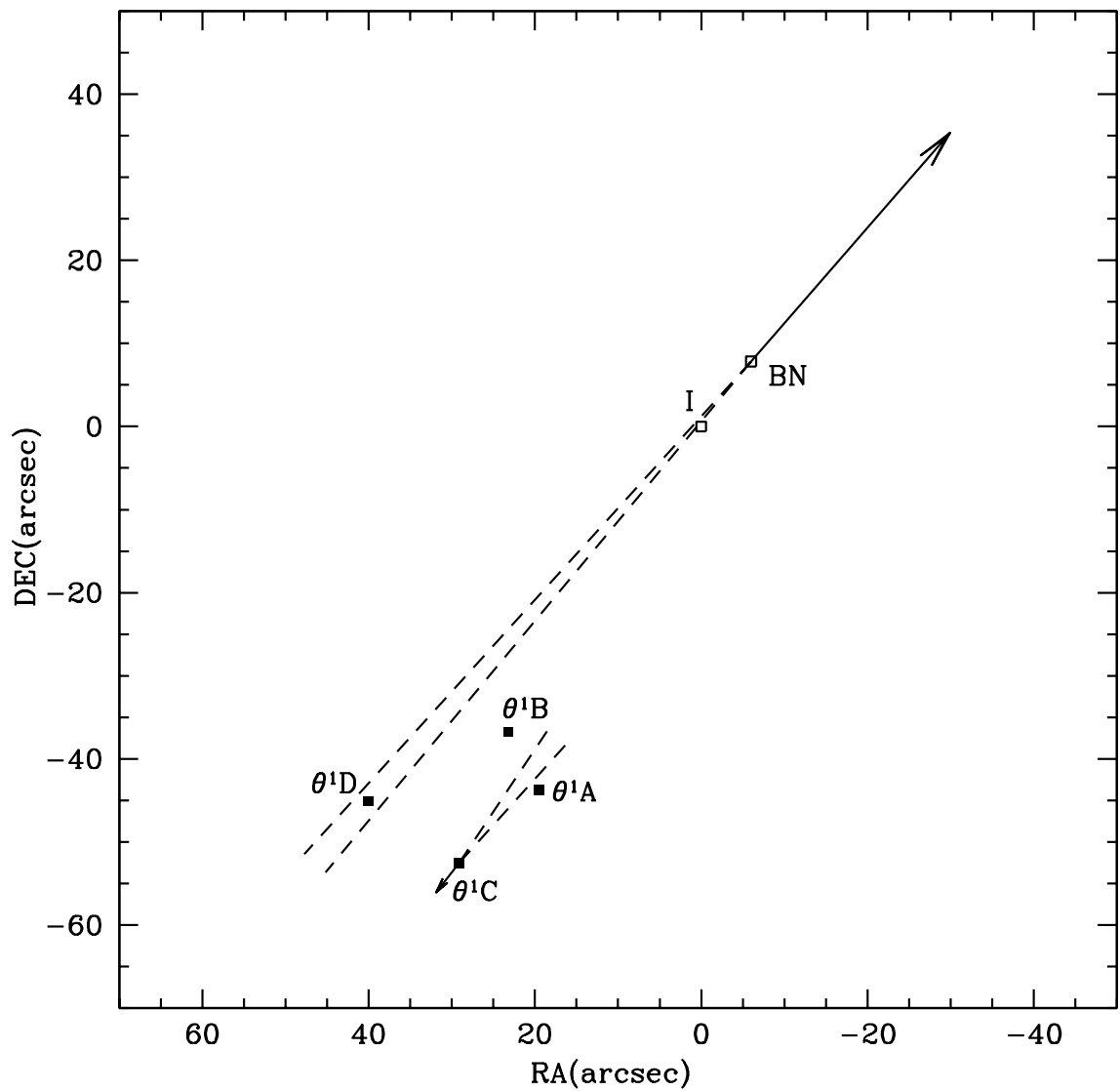}
\caption{The Orion BN/KL and Trapezium region is
shown in this figure. The sources BN and I are
indicated (empty squares), as well as the componens A, B, C, and D
of the Trapezium (filled squares). The proper motion of BN with respect to source I
is shown with an arrow that starts in BN and points
to the NW. 
The dashed lines that start in BN and point to the
SE indicate the past cone of uncertainty of the motions of BN.
The arrow and dashed lines that start in $\theta^1$ C Ori
mark the proper motion and past cone of uncertainty of the motions of
this source, respectively, taken from van Altena et al. (1988). 
In the arrows, one arcsec represents 0.5 mas yr$^{-1}$.
\label{fig6}}
\end{figure}

\clearpage

\begin{figure}
\epsscale{1.0}
\plotone{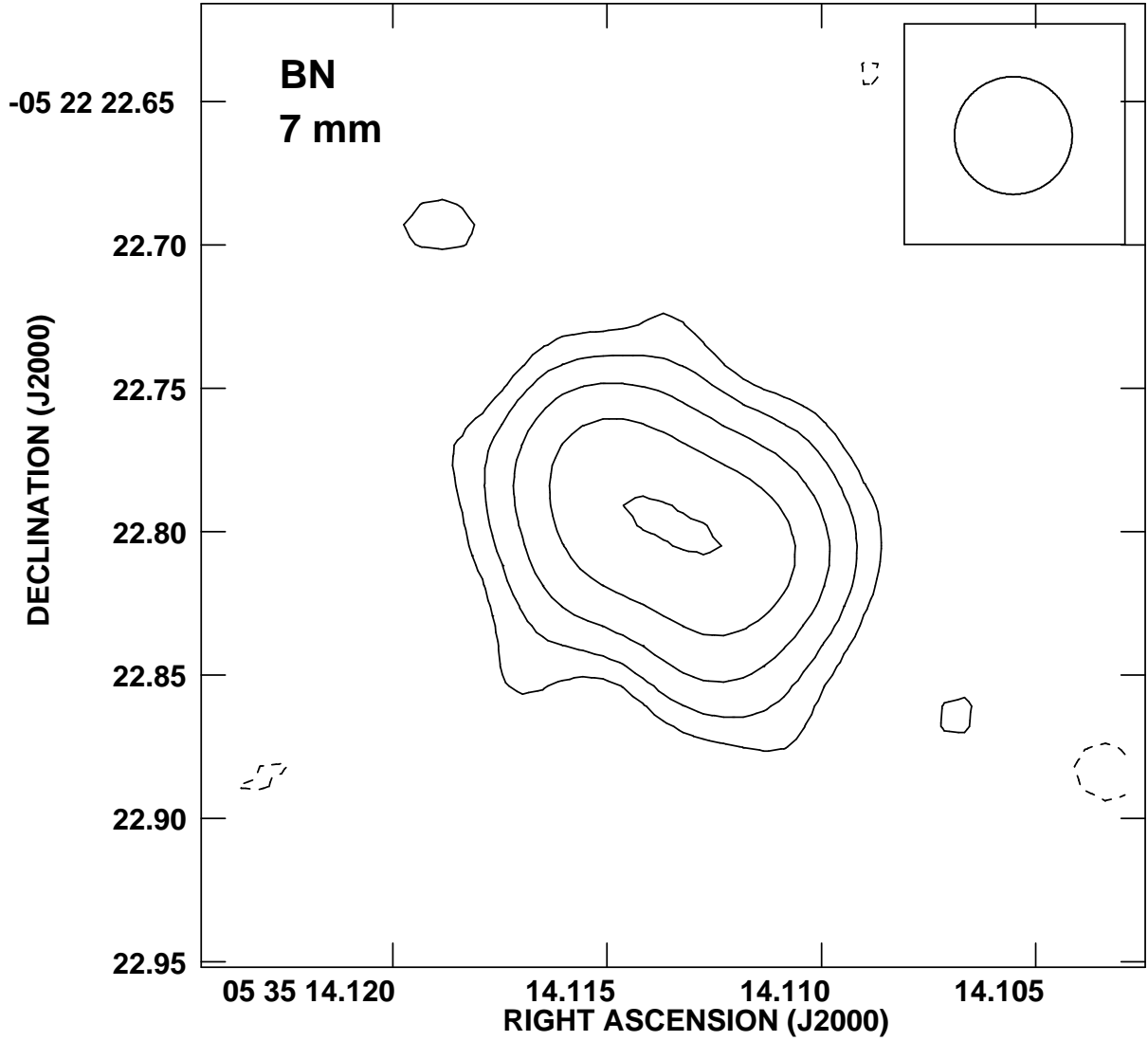}
\caption{VLA contour image at 7 mm
of BN, made from the dataset of Reid et al. (2007).
Contours are -3, 3, 6, 12, 24, and 48
times the rms noise of the image (0.17 mJy beam$^{-1}$).
The half power contour of the restoring beam, with diameter of
$0\rlap.{''}041$, is shown in the top right corner.
\label{fig7}}
\end{figure}

\clearpage

\begin{figure}
\epsscale{1.0}
\plotone{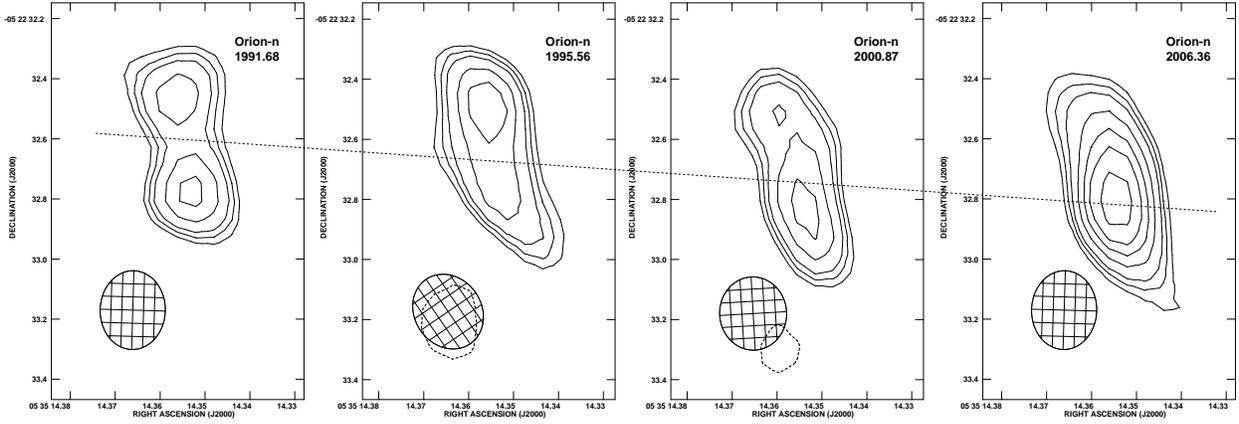}
\caption{VLA contour images at 8.46 GHz
of the source Orion-n for the four epochs discussed.
Contours are -4, 4, 5, 6, 8, 10 ,12 and 15
times the rms noise of the images
(99, 90, 55, and 58 $\mu$Jy beam$^{-1}$ for the 1991.68,
1995.56, 2000.87, and 2006.36 epochs, respectively).
The half power contour of the synthesized beams
($0\rlap.{''}26 \times 0\rlap.{''}25$; PA = $-54^\circ$,
$0\rlap.{''}26 \times 0\rlap.{''}22$; PA = $+34^\circ$,
$0\rlap.{''}24 \times 0\rlap.{''}22$; PA = $+3^\circ$, and
$0\rlap.{''}26 \times 0\rlap.{''}22$; PA = $-2^\circ$,
for the 1991.68,
1995.56, 2000.87, and 2006.36 epochs, respectively)
are shown in the bottom left corner of each panel
as a filled ellipse. The dashed line is intended
to guide the eye for the approximate proper motions over the years.
\label{fig8}}
\end{figure}

\clearpage

\begin{figure}
\epsscale{0.6}
\plotone{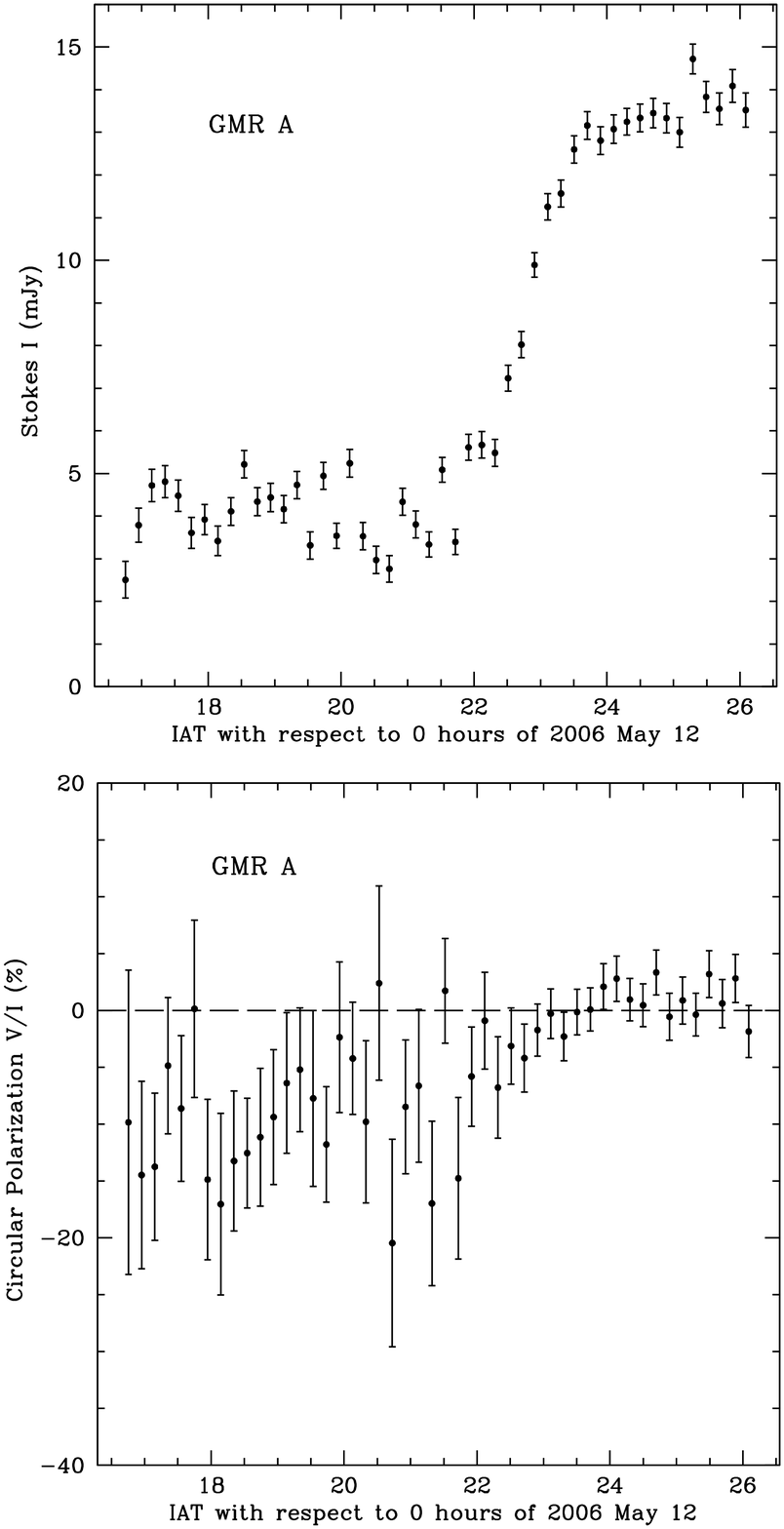}
\caption{Stokes I (top) and percentage of
circular polarization (bottom) for the source GMR A
during the 2006 May 12 flare.
\label{fig9}}
\end{figure}

\end{document}